\documentclass[twocolumn,pra,longbibliography,superscriptaddress]{revtex4-2} 
\usepackage{amsthm,amsmath,amssymb,amsfonts,graphicx,verbatim,xcolor,bm} 
\usepackage{ulem,hyperref} 
\usepackage[utf8]{inputenc} 
\usepackage{siunitx} 

\begin{document}

\title{Topological superconductivity in doped magnetic moir\'e semiconductors}
\author{Valentin Cr\'epel} 
\thanks{These authors contributed equally.}
\affiliation{Center for Computational Quantum Physics, Flatiron Institute, New York, New York 10010, USA}
\author{Daniele Guerci} 
\thanks{These authors contributed equally.}
\affiliation{Center for Computational Quantum Physics, Flatiron Institute, New York, New York 10010, USA}
\author{Jennifer Cano}
\affiliation{Department of Physics and Astronomy, Stony Brook University, Stony Brook, New York 11794, USA}
\affiliation{Center for Computational Quantum Physics, Flatiron Institute, New York, New York 10010, USA}
\author{J. H. Pixley}
\affiliation{Department of Physics and Astronomy, Center for Materials Theory, Rutgers University, Piscataway, New Jersey 08854, USA}
\affiliation{Center for Computational Quantum Physics, Flatiron Institute, New York, New York 10010, USA}
\author{Andrew Millis} 
\affiliation{Department of Physics, Columbia University, New York, NY 10027, USA} 
\affiliation{Center for Computational Quantum Physics, Flatiron Institute, New York, New York 10010, USA}

\begin{abstract}
We show that topological superconductivity may emerge upon doping of transition metal dichalcogenide heterobilayers above an integer-filling magnetic state of the topmost valence moir\'e band. The effective attraction between charge carriers is generated by an electric $p$-wave Feshbach resonance arising from interlayer excitonic physics and has a tuanble strength, which may be large. Together with the low moir\'e carrier densities reachable by gating, this robust attraction enables access to the long-sought $p$-wave BEC-BCS transition. The topological protection arises from an emergent time reversal symmetry occurring when the magnetic order and long wavelength magnetic fluctuations do not couple different valleys. The resulting topological superconductor features helical Majorana edge modes, leading to half-integer quantized spin-thermal Hall conductivity and to charge currents induced by circularly polarized light or other time-reversal symmetry-breaking fields.
\end{abstract}

\maketitle

\paragraph*{Introduction ---} Topological $p$-wave superconductors have been intensively sought after, in part because they host Majorana boundary modes~\cite{qi2011topological,leijnse2012introduction,flensberg2021engineered,crepel2019variational}, a key ingredient for the realization of topological quantum computations~\cite{nayak2008non}. Despite intensive efforts ~\cite{sato2017topological,sharma2022comprehensive}, topological superconductivity (TS) remains elusive, and material candidates are largely limited to fine-tuned non-stochiometric compounds~\cite{hor2010superconductivity,fu2010odd,xu2016topological,zhang2018observation,wang2018evidence,zhang2019multiple} that inevitably suffer from defects. The recent advent of gate-tunable moir\'e heterostructures allows one to bypass this unfavorable condition, and at the same time offers a unique context for intertwined topology and superconductivity~\cite{cao2018unconventional,park2022robust,balents2020superconductivity,andrei2021marvels,mak2022semiconductor,wu2019topological,crepel2022anomalous,devakul2021magic}.

Here, we propose a clear route towards the realization of helical $p$-wave superconductivity in transition metal dichalcogenide (TMD) moir\'e heterobilayers. This TS displays a valley-chirality locked $p \pm ip$ order protected by an emergent $\mathbb{Z}_2$ time-reversal symmetry. The TS arises in the weakly doped layer-transfer regime; where one layer contains one hole per moir\'e unit cell, forming a Mott insulator with a large gap to in-layer charge excitations, and an additional $x\ll 1$ carriers are added to the other layer, forming a dilute Fermi liquid. This regime has recently been reached in MoTe$_2$/WSe$_2$ heterobilayers~\cite{zhao2022gate}, where coexistence of local moments and itinerant carriers was reported~\cite{kumar2022gate,dalal2021orbitally,guerci2022chiral}.

\begin{figure}
\centering
\includegraphics[width=\columnwidth]{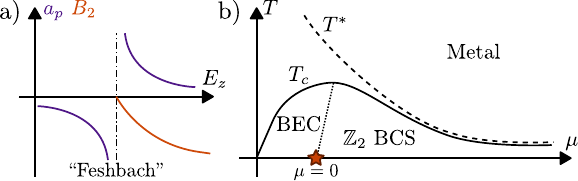}
\caption{a) An out-of-plane electric field $E_z$ can change the sign of $a_p$, the $p$-wave scattering length between doped charge. This solid-state Feshbach resonance yields bound states of energy $B_2<0$ below the band bottom when $a_p>0$~\cite{randeria1990superconductivity,zhang2017strongly}. b) Schematic phase diagram predicted for AA-stacked TMD heterobilayers as a function of chemical potential $\mu$, which can be varied by electrostatic gating. Below the BKT temperature $T_c$, the system either evolves into a $\mathbb{Z}_2$ topological superconductor ($\mu \gtrsim 0$) or a gapped Bose-Einstein condensate of pairs ($\mu \lesssim 0$). They are separated by a crossover at finite-temperature (dots), which becomes a phase transition at $T=0$ (star)~\cite{senthil2000z2,nussinov2008autocorrelations,qi2009time}. For large $\mu$, $T_c$ almost matches the pair binding $T^*$ (dashes).
}
\label{fig:SchematicPhaseDiag}
\end{figure}

For the physics addressed in this letter, the crucial feature of the TMD moir\'e bilayer is their strong interlayer Coulomb interaction, which leads to a remarkable range of (charged) interlayer excitons that strongly couple to mobile carriers. 
For example, full experimental control over electron-exciton scattering in TMD bilayers was recently demonstrated through interlayer trion dressing~\cite{schwartz2021electrically}, whose electric field dependent energy enabled scanning across a Feshbach resonance for this composite system~\cite{chin2010feshbach,kuhlenkamp2022tunable} 
In our theory, the crucial role is played by a low-lying charge-$2e$ interlayer exciton (quaternion), also recently probed by spectroscopy in TMD bilayers~\cite{sun2021charged}. This low-lying virtual state's contribution to electron scattering may be described in terms of an effective $p$-wave scattering length $a_p$ between doped charges that changes sign under electrostatic gating (see Fig.~\ref{fig:SchematicPhaseDiag}a). Proximity to this Feshbach resonance provides the strong interaction necessary for the emergence of robust superconductivity. 
Independent of superconductivity it is important to note that this $p$-wave electric Feshbach resonance is a solid-state realization of a phenomenon that is still intensely looked for in ultracold gases~\cite{regal2003tuning,zhang2004p,gunter2005p,gaebler2007p,top2021spin,park2023feshbach}. 

The physics of the weakly-doped layer-transfer regime is rich, with many different phases and phenomena depending on the stacking, interlayer potential and hybridization configurations. In this paper we focus on the arrangement of most interest for topological superconductivity, namely, AA-stacked bilayers with interlayer hybridization comparable to but weaker than the interlayer potential difference, which is in turn weaker than the interaction scales. The AB-stacked case and more general parameter regimes will be presented elsewhere~\cite{AFMcase}. In the case targeted here, the layer hybridization couples the itinerant carriers in the lightly doped layer to excitons, providing the $p$-wave pairing, and also leads to ferromagnetic order in the Mott insulating layer. This ferromagnetic order does not couple the Fermi pockets of the lightly doped layer, implying an emergent time reversal symmetry that promotes the $p$-wave superconducting state to the topologically protected DIII class~\cite{altland1997nonstandard}, which features helical Majorana edge modes.

Because the pairing depends on parameters independent of the Fermi surface and remains strong even in the very low density limit, we expect that the low carrier densities reachable by gating in moir\'e heterostructures enables access to the full evolution from weakly bound Cooper pairs forming a $\mathbb{Z}_2$ topological superconductor ($\mathbb{Z}_2$ BCS regime) to a Bose-Einstein condensate of tightly bound pairs (BEC regime), as sketched in Fig.~\ref{fig:SchematicPhaseDiag}b.

\begin{figure}
\centering
\includegraphics[width=\columnwidth]{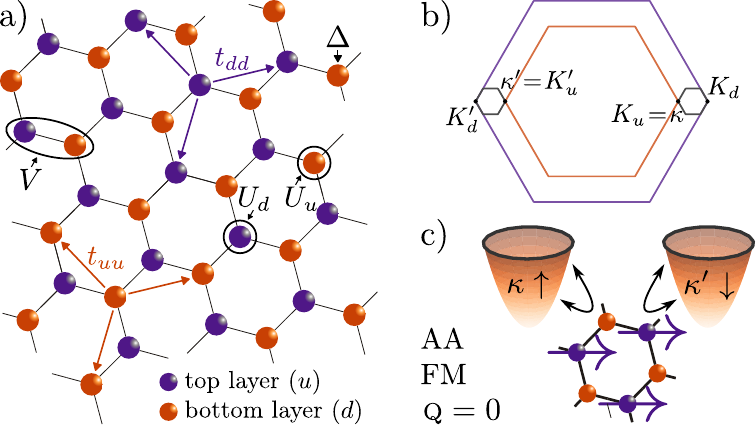}\vspace{-0.2cm}
\caption{a) Wannierized model of TMD heterobilayers (Eqs.~\ref{eq:Hab}\&\ref{eq:Hint}) keeping the dominant intra- and inter-layer interactions ($U_d$, $U_u$ and $V$) and tunnelings. The phases $\varphi_{uu}=\varphi_{dd}=2\pi/3$ of the tunnelings are depicted by arrows for the $K$ spin-valley component. 
b) Folding of the monolayer $\pm K_{u/d}$ points onto the mini-Brillouin zone corners $\kappa$ and $\kappa'$. 
c) In AA-stacked bilayers, the magnetic state stabilized at filling $n_h=1$ is a ferromagnet that cannot induce low-energy spin-flips due to spin-valley locking of the charge carriers described by parabolic dispersion around $\kappa$ and $\kappa'$.
}
\label{fig:ModelAndCoupling}
\end{figure}

\paragraph*{Model ---} Lattice parameter mismatch means that stacking two inequivalent TMD layers at zero or non-zero twist angle will create a moir\'e pattern with a unit cell that is large relative to atomic dimensions.  Extensive experimental~\cite{pizzocchero2016hot,mak2022semiconductor} and theoretical~\cite{zhang2016systematic,carr2018relaxation,massatt2021electronic,zhang2021spin,rademaker2022spin,pan2022topological} studies have established that the low energy physics of this situation may be described by a generalized Hubbard model $H=H_{\rm int}+H_{uu}+H_{dd}+H_{ud}$ involving two interpenetrating triangular lattices (one for each layer) featuring in-layer $H_{uu}, H_{dd}$ and interlayer $H_{ud}$ nearest neighbor hoppings~\cite{rademaker2022spin}, in-layer interactions $U_u, U_d$ and an interlayer interaction $V$ (see Fig.~\ref{fig:ModelAndCoupling}a). The hopping terms are
\begin{equation} \label{eq:Hab}
H_{ab} = -t_{ab} \sum_{\langle i ,j  \rangle_{ab}} c_i^\dagger e^{-i \sigma^z \nu_{ij}^{ab} \varphi_{ab}} c_j ,
\end{equation}  
with $(a,b) \in \{ u,d \}$ denoting the up or down layer, $(i,j)$ labelling orbitals in these layers, and $\langle i , j \rangle_{ab}$ denoting nearest neighbor pairs having $i \in a$ and $j \in b$ (see Fig.~\ref{fig:ModelAndCoupling}a). We have written the fermionic operator for holes as two-component spinors $c_{j}^\dagger= [ c_{j,K}^\dagger, c_{j,K'}^\dagger ]$, labelled by the  spin-valley locked degrees of freedom of the two wannierized TMDs. The hopping parameters are in general complex~\cite{wu2019topological}; in the `AA' stacked configuration studied here, we may choose the interlayer hopping parameters to be real ($\varphi_{ud}=0$) and set $\varphi_{uu} = \varphi_{dd} = 2\pi/3$ with $\nu_{ij} = + 1$ when the link $i\to j$ turns right and $\nu_{ij} = - 1$ otherwise. In this convention, the $t_{ab}$ are real and positive. The interaction terms may be written
\begin{equation} \label{eq:Hint}
H_{\rm int} = \Delta \sum_{i \in u} n_i + \sum_{a, i\in a} U_a n_{i,\uparrow} n_{i,\downarrow} + V \sum_{\langle i , j \rangle_{ud} } n_i n_j .
\end{equation}
The interlayer potential difference $\Delta$ is about $\SI{0.1}{\electronvolt}$ for the MoTe$_2$/WSe$_2$ system of immediate experimental relevance, and may be tuned by an out-of-plane electric field~\cite{li2021quantum,zhao2022realization}.

Representative values estimated from a continuum model for MoTe$_2$/WSe$_2$ bilayers~\cite{zhang2021spin} are $U_{d}\approx \SI{0.30}{\electronvolt} \gtrsim \Delta$, $U_{u}\approx\SI{0.24}{\electronvolt}$ and $V\approx\SI{0.14}{\electronvolt}$, the large size of the moir\'e unit cell relative to the interplane separation explaining $V\sim U_{u,d}$. The superconducting state discussed in this work appears when $V$ exceeds $\Delta/4$ (see below). The interaction strength increases with increasing $V$ until $V$ becomes so large that the charge transfer gap exceeds the Mott gap and the doped holes go into the magnetic layer.

\paragraph*{Magnetic coupling ---} We investigate the physics of the Hamiltonian defined by Eqs.~\ref{eq:Hab}\&\ref{eq:Hint} in the layer-transfer limit $t \equiv t_{ud} \ll \Delta \ll \Delta+3V<U_{d}$. In this regime, carriers added up to a density of one hole per per moir\'e unit cell go into the lower layer and due to the large $U_d$ form a Mott insulator at the density $n_h=1$, while a small density $x$ of carriers added beyond $n_h=1$ will go into the upper plane. We now consider the interactions affecting these $x$ extra carriers.

When $U_{d}\gg\Delta$, the leading magnetic interaction is a trion-mediated exchange which, combined with the strong single-layer spin orbit coupling, leads to $xy$-ferromagnetism in the Mott layer~\cite{devakul2021magic}. In-plane magnetism in the lower level acts as a spin-flip operator for carriers in the upper layer. However, the spin-valley locking in the monolayers, transferred to the moir\'e $\pm \kappa$ valleys after downfolding (see Fig.~\ref{fig:ModelAndCoupling}b), means that low energy spin-flips involve momentum transfers of the order of $\kappa-\kappa^\prime$. For this reason, a small density of carriers doped above the Mott insulating state cannot undergo spin-flip scattering from the ferromagnetic order or its low-lying spin wave excitations at low-energy (see Fig.~\ref{fig:ModelAndCoupling}c).  As a result, the bottom layer effectively behaves as a featureless charge reservoir and the low-energy carriers in AA-stacked bilayers enjoy both an emergent time reversal symmetry (TRS) $T=i\tilde\sigma^y K$ and the full U(1) spin-rotation symmetry generated by $\tilde \sigma^z$, although both are spontaneously broken by the Mott state. Here, $\tilde\sigma$ denotes the spin-valley Pauli matrices projected to the active modes $[\psi_{q,\Uparrow}^\dagger, \psi_{q,\Downarrow}^\dagger]$ near the top of the valence band, with $\Uparrow/\Downarrow = (\kappa/\kappa^\prime, \uparrow/\downarrow)$. The system features two spin-valley locked hole pockets with dispersion $\varepsilon_q = q^2 / 2 m$~\cite{Suppmat}, shown in Fig.~\ref{fig:ModelAndCoupling}c, related by the emergent TRS.

\begin{figure}
\centering
\includegraphics[width=\columnwidth]{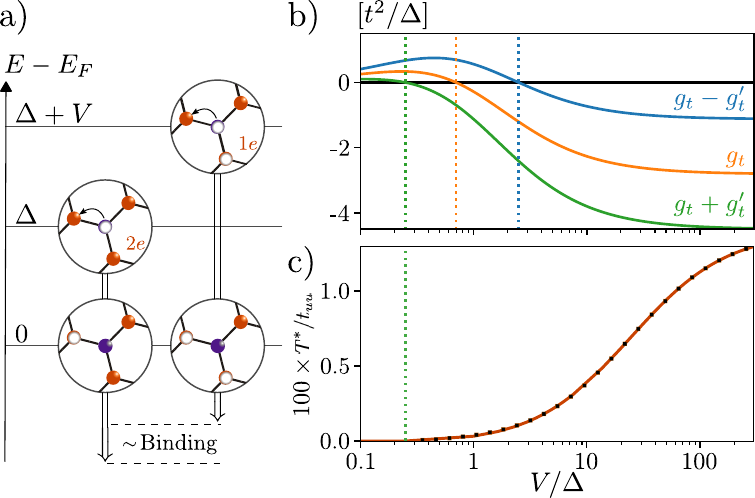}
\caption{
a) The lowest excitation with which isolated carriers (orange) hybridize has higher energy than the charge $2e$ excitons that couples to pairs of carriers. This offers a strong energy reduction to pairs and can produce a non-zero binding energy.
b) Interaction coefficients in the different $p$-wave channels (Eq.~\ref{eq_interactioncontinuum}) with dotted lines marking the value of $V/\Delta$ above which they become negative -- 1/4 for the leading pairing channel. 
c) The pair binding energy $T^*$ is non-zero above this value, and increases with $V/\Delta$ from small to large values compared to the Fermi energy $E_F$, describing a BCS to BEC evolution. The solid and dotted lines respectively show numerical estimates and results from a generic $p$-wave BCS formula, both obtained for $x=0.1$, $t_{uu}=\Delta$ and $t^2/(\Delta t_{uu})=0.25$. 
}
\label{fig:Pairing}
\end{figure}

\paragraph*{Equal-spin pairing instability ---} Since carriers near the Fermi surface (FS) only couple to the density of the insulating bottom layer, our system is an experimentally viable realization of the setup recently discussed in terms of model systems to describe a repulsive mechanism for superconductivity~\cite{Slagle20,crepel2021new,crepel2022spin,crepel2022unconventional,he2023superconductivity}.
This mechanism relies on the existence of a charge-$2e$ exciton (quaternion) with lower energy than all charge $e$ and neutral excitations of the system at $t=0$, which is achieved thanks to the large $V$ of our model. This quaternion provides a closed scattering channel that can be virtually occupied by pairs of mobile carriers to obtain a non-zero binding energy (see Fig.~\ref{fig:Pairing}a), in direct analogy to the physics of Feshbach resonance.

The effective attraction is explicitly seen when the interaction term of our lattice model is projected onto the active modes at low doping, retaining pair operators with relative form factor of zeroth or first order in the small momentum deviations away from $\pm \kappa$~\cite{Suppmat}
\begin{align} \label{eq_interactioncontinuum}
\mathcal{H}_{k,k'}^{\rm int} = g_s \mathcal{S}_k^\dagger \mathcal{S}_{k'} + \sum_{\ell s} (g_t - \ell s g_t' ) \left(\mathcal{T}_k^{\ell s}\right)^\dagger \mathcal{T}_{k'}^{\ell s} , 
\end{align}
where $\mathcal{S}_k = ( \psi_{- k, \Uparrow} \psi_{ k, \Downarrow} - \psi_{- k, \Downarrow} \psi_{ k, \Uparrow} ) / \sqrt{2}$ denotes the $s$-wave pair operator, while $\mathcal{T}_k^{\ell -} = k_\ell  \psi_{- k, \Downarrow} \psi_{ k, \Downarrow}$, $\mathcal{T}_k^{\ell 0}= k_\ell ( \psi_{- k, \Uparrow} \psi_{ k, \Downarrow} + \psi_{- k, \Downarrow} \psi_{ k, \Uparrow}) / \sqrt{2}$ and $\mathcal{T}_k^{\ell +} = k_\ell  \psi_{- k, \Uparrow} \psi_{ k, \Uparrow}$ describe $p$-wave pairs of spin $s=-1,0,+1$, respectively. Their orbital angular momentum $\ell = \pm$ is fixed by their form factors $k_{\pm}=k_x\pm ik_y$. As claimed, the $g$-coefficients extracted from second order perturbation theory~\cite{Suppmat}, plotted in Fig.~\ref{fig:Pairing}b, unveil attractive interactions in the $p$-wave channel of our model for large enough $V/\Delta$~\cite{Suppmat}. The $s$-wave scattering amplitude receives a contribution from the large on-site repulsion and therefore remains positive for our parameter regime, $g_s \sim U_u >0$. The largest negative interaction strength is found in the $\{ \mathcal{T}^{-+}, \mathcal{T}^{+-} \}$ sector, which describes valley-chirality locked $p \pm ip$ equal-spin pairing.

The pair binding energy $T^*$ extracted from the log-singularity of the particle-particle susceptibility in these channels is shown in Fig.~\ref{fig:Pairing}c~\cite{Suppmat}. It perfectly agrees with the generic BCS-like formula for $p$-wave attraction $k_B T_c \propto \exp[- 1/(\rho E_F \tilde{g})]$~\cite{maiti2013superconductivity}, where $\tilde{g} = 4 \rho |g_t + g_t'| / \pi$ is the dimensionless attraction strength in the dominant pairing channel, and $\rho = m / (2 \pi \hbar^2)$ the constant density of states near the band bottom~\footnote{We fixed the unit cell area to one}. Continuum model calculations give the gap-to-$T^*$ ratio $2\Delta_{\rm sc}/(k_BT^*)\approx 3$.


\paragraph*{$\mathbb{Z}_2$ topological superconductor ---} We now show that the emergent low-energy TRS of doped holes grants topological protection to the superconducting state, resulting in pairs of helical Majorana modes on its edges. Introducing the bosonic fields $\phi_\pm$ to describe the superconducting order parameters in the $\mathcal{T}^{\pm \mp}$ channels, and performing a Hubbard-Stratonovitch transformation, we obtain the Bodgoliubov-de Gennes (BdG) Hamiltonian 
\begin{equation} \label{eq_meanfieldBdG}
\mathcal{H}_q^{\rm BdG} = \frac{1}{2} \begin{bmatrix} h_q & \Delta_q \\ \Delta_q^\dagger & -h_q \end{bmatrix} , \quad \Delta_q = \begin{bmatrix} 0 & \phi_+ q_+ \\ \phi_- q_- & 0 \end{bmatrix} ,
\end{equation}
expressed in Nambu space $[\psi_{q,\Uparrow},\psi_{q,\Downarrow}, \psi_{-q,\Downarrow}^\dagger,\psi_{-q,\Uparrow}^\dagger]$, with $h_q = q^2/2m - \mu$ and $\mu$ the chemical potential. The block structure of $\mathcal{H}^{\rm BdG}$ translates into a decoupled sum of free energies for the $\Uparrow/\Downarrow$ sectors $\mathcal F =\sum_{a=\pm}( \alpha |\phi_a|^2 + |\phi_a|^4)$. Below $T_c$, \textit{i.e.} for $\alpha<0$, the minimization of $\mathcal{F}$ implies that both species be equally populated $|\phi_+|=|\phi_-|$~\cite{Suppmat}. Up to an irrelevant gauge choice, we thus have $\phi_+=\phi_-^*=\phi e^{i\theta}$.

\begin{figure}
\centering
\includegraphics[width=\columnwidth]{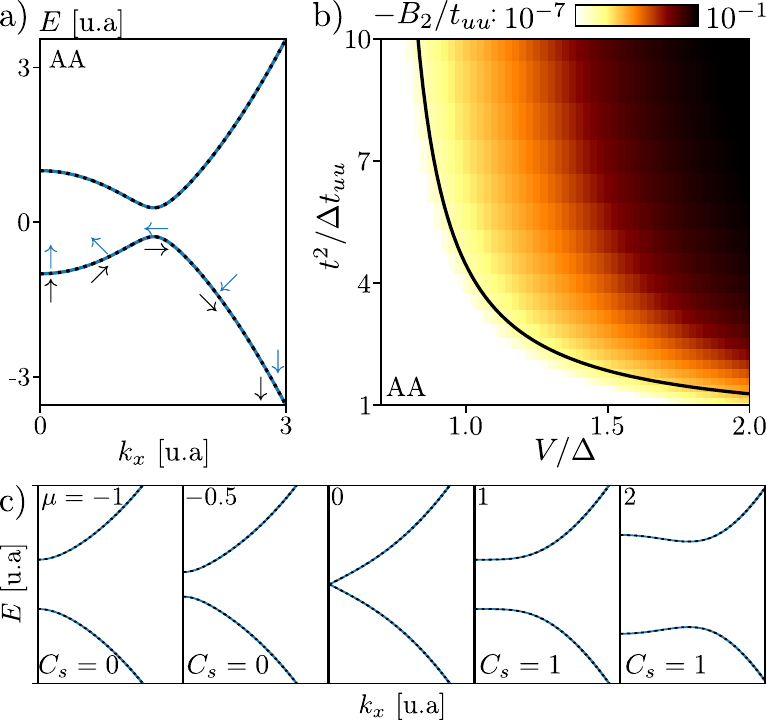}
\caption{a) BdG band structure for the superconducting state. The non-trivial and opposite winding of the Nambu vectors of the lower bands $n_{q}^s$, shown with arrows, signals a non-zero spin Chern number. 
b) Pairing persists down to the two-particle level for certain choices of parameters, as indicated by a negative binding energy $B_2<0$. The black lines show the separation between the regions with and without bound states below the carrier band edge obtained from the effective continuum theory Eq.~\ref{eq_interactioncontinuum}.
c) The evolution from BEC (left) to BCS (right) at zero temperature involves a topological phase transition, illustrated here using the BdG band structure as a function of the chemical potential $\mu$. $C_s$ denotes the spin Chern number of the negative-energy BdG bands.
}
\label{fig:Bdg}
\end{figure}

The two spin-valley components of $\mathcal{H}^{\rm BdG}$ decouple into time-reversal conjugated $2\times2$ blocks that can be written using Nambu Pauli matrices $\vec{\tau}$ as $\mathcal{H}_q^s = E_q \vec{n}_q^s  \cdot \vec{\tau}$ with $E_q^2 = h_q^2 + |\phi q|^2$. As illustrated in Fig.~\ref{fig:Bdg}a, the unit vectors $n_q^s = [  \phi \left(R_\theta q\right)_x , s \phi \left(R_\theta  q\right)_y , h_q ]/E_q$, where $R_\theta$ is the rotation matrix by angle $\theta$ around the $z$-axis, fully wrap around the Bloch sphere as momentum is varied provided $\mu >0$. This ensures a non-zero Chern number to all four Bogoliubov bands. Since the vectors $n_q^s$ are mirrors of one another with respect to the $(xz)$ plane for opposite spin $s=\pm$, the Chern numbers for the two hole-like Bogoliubov bands are opposite. They hence carry a non-trivial spin Chern number $C_s = 1$.

The existence of this spin-Chern number follows from the particle-hole operator $\mathcal{P} = \tilde\sigma^0 \tau^x$ acting as $\mathcal{P} \mathcal{H}_q^{\rm BdG} \mathcal{P} = -\mathcal{H}_{-q}^{\rm BdG}$ and an additional chiral symmetry $\mathcal{O}=\tilde\sigma^z \tau^z$ that commute with $\mathcal{H}_q^{\rm BdG}$ which imply that the state belongs to the DIII class of the Altland-Zirnbauer classification~\cite{altland1997nonstandard} and the obtained triplet superconducting state is topologically protected. The topological phase displays a superposition of $p\pm ip$ superconducting components~\cite{schnyder2008classification}, easily anticipated given the form of $\Delta_q$ in Eq.~\ref{eq_meanfieldBdG}. More remarkable is the existence of counter-propagating chiral Majorana modes at the edge of the system, described by
\begin{equation} \label{eq_Majorana}
\mathcal{L} = i \chi_\Uparrow (\partial_t - \partial_x) \chi_\Uparrow + i \chi_\Downarrow (\partial_t + \partial_x) \chi_\Downarrow ,
\end{equation}
where $\chi_\Uparrow = u\psi_{\Uparrow} + u^*\psi^\dagger_{\Uparrow}$ and $\chi_\Downarrow = u^*\psi_{\Downarrow} + u\psi^\dagger_{\Downarrow}$, with $u$ the normalized hole component of the $\Uparrow$ band of $\mathcal{H}_q^{\rm BdG}$~\cite{nayak2008non}. This pair of helical edge modes is protected by $\mathcal{O}$, for which they are eigenmodes with opposite eigenvalues.     

The obtained $\mathbb{Z}_2$ topological superconductor can be revealed by a half-integer value of the spin-thermal Hall conductivity, coming from the Majorana edge modes~\cite{read2000paired}. While spin-thermal Hall currents have not yet been measured, any TRS breaking perturbation, \textit{e.g.} circularly polarized light, can imbalance the population in the two valleys and confer properties akin to chiral $p$-wave superconductors, which can be probed through electric currents~\cite{kallin2016chiral}. In addition, Majorana zero modes in vortex cores~\cite{read2000paired} could be observed with scanning tunneling microscopy as zero-bias peaks~\cite{jack2021detecting}.


Beyond its topological protection, another interesting feature arises from the large momentum transfer $\kappa-\kappa^\prime$ required to coherently scatter holes between the two valleys, which prevents any quartic term besides the residual pair repulsion from appearing in the Ginzburg free energy~\cite{Suppmat}. The lowest order term coupling the phases of $\phi_+$ and $\phi_-$ is thus of sixth order $(\phi_+^{*})^3 \phi_-^3 +hc$. This high order term induces a noteworthy third-order Josephson effect with distinctive low-energy Leggett modes~\cite {leggett1966number}. 

\paragraph*{$p$-wave BEC-BCS transition ---} $T^*$ in Fig.~\ref{fig:Pairing}c only matches the BKT transition temperature $T_c$ in the weak-coupling regime $E_F \gg |g_t+\tilde g_t|$~\cite{maiti2013superconductivity}. In the opposite strong-coupling limit, \textit{i.e.} at doping concentrations $x = 2\rho E_F < \rho  |g_t+\tilde g_t|$, our model still exhibits pairing in some regime of parameters. To see this, we consider the binding energy of two charge carriers doped above the magnetic state $B_2 = E_2 - 2E_1 + E_0$, $E_N$ denoting the ground state energy for $N$ doped charges. 

We first obtain $B_2$ as a function of the original parameters of the model by solving an effective lattice model containing all second order processes in $t/\Delta$~\cite{Suppmat}. The results of this calculation are presented in Fig.~\ref{fig:Bdg}b, and show the following trend: as $\Delta$ is decreased, \textit{e.g.} by application of an out-of-plane electric field, the ratios $V/\Delta$ and $t^2/t_{uu} \Delta$ increase up to a critical point where bound states emerge $B_2<0$. This can be seen as a condensed-matter analog of a Feshbach resonance, where the non-retarded interaction between two fermions can be tuned from positive to negative using an externally controllable parameter. From a low-energy scattering perspective, this can be understood as tuning the $p$-wave scattering length $a_p$ from negative to positive. The universal relation $B_2 \sim \hbar^2 / m a_p^2 \log ( r / a_p )$ holds when $a_p>0$~\cite{randeria1990superconductivity,zhang2017strongly}, where $r$ is the range of $p$-wave interactions, comparable to the lattice constant.

The presence of bound pairs at infinitesimal doping offers access to the full evolution from a BEC of pairs to the BCS superconducting state, studied above in the weak coupling limit (Fig.~\ref{fig:Pairing}). This evolution should be distinguished from the $s$-wave case in several ways. For $p$-wave interactions, the BCS and BEC regions are separated by a transition, \textit{i.e.} by a gap closing~\cite{gurarie2005quantum}, while it is a smooth crossover for $s$-wave interactions~\cite{randeria2014crossover}. This is easily observed in our BdG Hamiltonian, whose eigen-energies vanish at $\mu = q = 0$ even when $\phi >0$ remains finite. This gap closure, highlighted in Fig.~\ref{fig:Bdg}c, separates the BCS regime $\mu > 0$ from the BEC regime $\mu<0$. Another difference is that the physics in the $p$-wave case necessarily involves another length-scale in addition to the scattering length~\cite{gurarie2005quantum,gurarie2007resonantly}.

In our specific model, the topological protection of the superconducting state in the FM case endows the BEC-BCS transition with a topological character. This is understood from the spin-split BdG Hamiltonians $\mathcal{H}_q^s$, which exhibit a textbook example of a band-inversion when the ``mass'' $\mu$ crosses zero energy~\cite{hasan2010colloquium}. The topological properties of the superconducting state are lost at any finite temperature due to thermal proliferation of topological excitations~\cite{senthil2000z2,nussinov2008autocorrelations,qi2009time}, and as a result the $T=0$ transition into a crossover at any finite temperature.




\paragraph*{Conclusion ---} We have exposed physical mechanisms leading to the emergence of a robust attraction and a low-energy time-reversal symmetry in AA-stacked transition metal dichalcogenides moir\'e heterobilayers doped above unit filling, which together produce a topologically protected helical $p$-wave superconductor at sufficiently low temperatures. The topological properties are inherited from the strong spin-orbit coupling of the original monolayers, when the latter is preserved by the magnetic Mott state. The attraction relies on the large interlayer interaction $V$ of the bilayer, and increases when electrostatic gating reduces the valence band offset $\Delta$ between layers. For our theory to apply, the layer-transfer gap should also remain smaller than the in-layer Mott gap at filling one. All these scales can, in principle, be experimentally probed by scanning-tunneling microscopy or compressibility measurements to provide experimental guidance on how to reach the regime of interest for superconductivity.

\paragraph*{Acknowledgments ---}

D.G. thanks Michele Fabrizio for correspondence at the early stage of the work. V.C. is grateful to A. Imamoglu for an insightful discussion shaping some of the ideas presented here. We also acknowledge enlightening discussions with Chetan Nayak. 
This work was partially supported by the Air Force Office of Scientific Research under Grant No.~FA9550-20-1-0260 (J.C.) and Grant No.~FA9550-20-1-0136 (J.H.P.) and the Alfred P. Sloan Foundation through a Sloan Research Fellowship (J.C., J.H.P.). A.J. M. acknowledges support from the NSF MRSEC program through the Center for Precision-Assembled Quantum Materials (PAQM)
NSF-DMR-2011738.
The Flatiron Institute is a division of the Simons Foundation.

\bibliography{BibKondoAttraction}


\onecolumngrid
\newpage
\makeatletter 

\begin{center}
\textbf{\large Supplementary material for: `` \@title ''} \\[10pt]
Valentin Cr\'epel$^1$
, Daniele Guerci$^1$, Jennifer Cano$^{2,1}$, J. H. Pixley$^{3,1}$, Andrew Millis$^{4,1}$ \\
\textit{$^1$ Center for Computational Quantum Physics, Flatiron Institute, New York, New York 10010, USA}\\
\textit{$^2$ Department of Physics and Astronomy, Stony Brook University, Stony Brook, New York 11794, USA}\\
\textit{$^3$ Department of Physics and Astronomy, Center for Materials Theory, Rutgers University, Piscataway, New Jersey 08854, USA}\\
\textit{$^4$ Department of Physics, Columbia University, New York, NY 10027, USA}
\end{center}
\vspace{20pt}

\setcounter{figure}{0}
\setcounter{section}{0}
\setcounter{equation}{0}

\renewcommand{\thefigure}{S\@arabic\c@figure}
\makeatother

\appendix


These supplementary materials contain the details of analytic calculations as well as additional numerical details supporting the results presented in the main text. It is structured as follows. In Sec.~\ref{app:SW_transformation} we discuss the  Schrieffer-Wolff transformation 
providing the effective lattice model describing the doped charges' dynamics, 
and expand it in the low density limit, 
which gives rise to the low-energy theory presented in the main text. In Sec.~\ref{app:two_doped} and~\ref{app:Tc} we discuss the 2$e$ bound states forming the dilute BEC and the expression for $T^*$ in the BCS regime, respectively. Finally, in Sec.~\ref{app:density_density} we comment on a quartic symmetry allowed term coupling the density $|\phi_+|$ and $|\phi_-|$ of the two spin-valley locked condensates.  

\section{Schrieffer-Wolff transformation}
\label{app:SW_transformation}


In this section we detail the effective Hamiltonian of itinerant carriers hybridizing with the ferromagnetically ordered moments. We perform an expansion in the interlayer parameter $t_{ud}$, which we recall is small compared to the charge transfer gap $\Delta$.

\subsection{Effective lattice model}
\label{app:effective_model}

Accounting for all possible virtual processes we find that the lattice Hamiltonian for the doped charges reads
\begin{equation}  \begin{split}
\label{eq_effectivemodelinfiniteU}
H_{\rm eff} = &-t_{uu} \sum_{\langle i ,j  \rangle_{ab}} c_i^\dagger e^{-i \sigma^z \nu_{ij}^{uu} \varphi} c_j + U_u\sum_i n_{i\uparrow} n_{i\downarrow} \\
&+ \sum^{i\neq j\neq k}_{ijk \in \triangle} \left\{ c_{i}^\dagger \left[ \bm{S}_A \cdot \bm{\sigma} + \frac{1}{2} \right] (t_K + \lambda n_k) c_j  + P_{ijk}\right\} + \left[ U_2 + U_3 \frac{n_\triangle-2}{3} \right] \frac{n_\triangle(n_\triangle-1)}{2}  \, ,
\end{split}
\end{equation}
where $n_\triangle = n_i +n_j+n_k$ is the total density on the lower triangles $\triangle$ forming the $u$ lattice, $P_{ijk}$ denotes summation over all permutation of the indices $ijk$, and $[\bm{S}_A \cdot \bm{\sigma}+1/2]$ describes the exchange of spin between the $u$ carriers and the nearest $d$ spin. Finally, we replace the spin operators $\bm{S}_A$ by their classical values $\bm{S}_A\to \langle \bm{S}_A\rangle$ assuming long-range magnetic order in the $d$ layer and neglecting low-energy spin wave excitations.
Setting $t=t_{ud}$, the coefficients appearing in $H_{\rm eff}$~\ref{eq_effectivemodelinfiniteU} are
\begin{equation} 
\label{pt_coefficients}
t_K = \frac{t^2 }{\Delta+V} , \quad \lambda = \frac{t^2 }{\Delta} - \frac{t^2 }{\Delta +V} , \quad
U_2 = - \frac{t^2 }{\Delta} + \frac{4 t^2 }{\Delta+V} - \frac{3t^2 }{\Delta + 2V} , \quad
U_3 =  \frac{3t^2 }{\Delta} - \frac{6 t^2 }{\Delta+V} + \frac{3t^2 }{\Delta + 2V} .
\end{equation} 

\subsection{Low-energy projection}
\label{app:low_energy}

At low density of doped carriers in the $u$ layer we expand Eq.~\ref{eq_effectivemodelinfiniteU} close to the bottom of the itinerant band by retaining only the spin-valley locked fermionic modes at a distance $|q| \ll 1$ from the $\pm \kappa$ points, gathered in the spinor $\tilde\Psi_{q}=(\psi_{q,\Uparrow},\psi_{q,\Downarrow})^T$ with $\Uparrow/\Downarrow = (\kappa/\kappa',\uparrow/\downarrow)$. The kinetic part of the Hamiltonian is simply $q^2/2m$ with $m^{-1}=9(t_K+4t_{uu})/4$. Moreover, the interaction coefficients in Eq.~\ref{eq_interactioncontinuum} of the manuscript read $g_s = U + 6U_2-3\lambda$, $g_t = 9(\lambda - 4U_2)/8$, $g_t' = 27 \lambda /8$, where $\lambda$ and $U_2$ are given in Eq.~\ref{pt_coefficients}.

\section{Two doped charges}
\label{app:two_doped}

The binding energy $B_2$ shown in Fig.~\ref{fig:Bdg}b are obtained by solving the effective tight-binding model Eq.~\ref{eq_effectivemodelinfiniteU} for zero, one and two particles. In the last case, we decouple the center of mass and the relative motion to access larger system sizes. 

The solid black line in Fig.~\ref{fig:Bdg}b showing the separation between the regions with and without bound states is obtained looking for the poles of the T matrix~\cite{levinsen2015strongly}:
\begin{equation}
    -\frac{1}{g_t+ g_t'}=\Pi_t(-|\epsilon|),
\end{equation}
where the polarization bubble takes the form:
\begin{equation}
   \Pi_t(-|\epsilon|)=\int^\Lambda\frac{d^2k}{(2\pi)^2}  \frac{k^2}{|\epsilon|+k^2/2m},
\end{equation}
where
$\epsilon=-|\epsilon|<0$ is negative and $\Lambda$ is the UV cutoff. The value of $\Lambda$ is obtained by fitting the numerical results from the lattice model.

\section{Pair binding energy}
\label{app:Tc}

Our weak coupling estimate for $T^*$ is obtained by solving
\begin{equation}
1 = - [ g_t+ g_t'] \chi_t(T^*), 
\end{equation}
where the triplet susceptibility reads 
\begin{equation}
\chi_{t} (T) = \int^\Lambda\frac{d^2k}{2\pi^2}   \frac{k^2}{k^2/2m-\mu}\tanh\frac{k^2/2m-\mu}{2T}.
\end{equation}
In the latter expression the chemical potential is fixed by:
\begin{equation}
x = \int^\Lambda\frac{d^2k}{2\pi^2}  f\left(\frac{k^2}{2m}-\mu\right).
\end{equation}
with $f$ the Fermi-Dirac distribution. Finally, we also give the expression for the $T=0$ superconducting gap equation: 
\begin{equation}
    1=-(g_t+g'_t)\int^\Lambda\frac{d^2k}{4\pi^2}\frac{k^2}{\sqrt{(k^2/2m-\mu)^2+|\phi|^2k^2}},
\end{equation}
with $|\phi|=|\phi^+_+|=|\phi^-_-|$.

\section{Symmetry allowed quartic term in GL free energy}
\label{app:density_density}

Including longer range interactions can mix the two order parameters $\phi_+$ and $\phi_-$ at the quartic level through a correction of the Ginzburg-Landau free energy $\delta \mathcal{F} = \beta |\phi_+ \phi_-|^2$. Alll other quartic terms mixing the two superconducting orders involve large momentum transfer $\pm \kappa$ and thus average to zero. Including the dominant intra- and inter- layer interactions yields $\beta = 0$ (see main text), and we expect that the sub-dominant interactions discarded in our model Eqs.~\ref{eq:Hab}\&\ref{eq:Hint} will keep the $\beta$ coefficient small compared to one. In particular, so long as $\beta < 2$, we will continue to have $|\phi_+|=|\phi_-|$ and all the results derived in the main text remain valid.

\end{document}